\documentclass[reprint,prl,showpacs]{revtex4-1}
\usepackage{bm}
\usepackage{graphicx}
\usepackage{amsmath}
\usepackage{hyperref}
\usepackage{txfonts}

\begin{document}

\title{Symmetry Protected Quantization 
and Bulk--Edge Correspondence of Massless Dirac Fermions:
 Application to Fermionic Shastry-Sutherland Model}

\date{\today}
\author{Toshikaze Kariyado}\email{kariyado.t.gf@u.tsukuba.ac.jp}
\author{Yasuhiro Hatsugai}
\affiliation{Division of Physics, Faculty of Pure and Applied Sciences,
University of Tsukuba, Tsukuba, Ibaraki 305-8571, Japan}

\begin{abstract}
The fermionic Shastry-Sutherland model has a 
 rich phase diagram, including phases
 with massless Dirac fermions, a quadratic band crossing point, and a
 pseudospin-1 Weyl fermion. 
Berry phases 
defined by the one-dimensional momentum as a parameter
are quantized into $0$ or $\pi$ 
due to the inversion symmetry combined with the time reversal, or
existence of the glide plane,
which also protects the massless Dirac cones 
with continuous parameters. 
This is the symmetry protected $Z_2$ quantization. 
We have further 
 demonstrated the $Z_2$ Berry phases generically 
determine the existence of 
edge states in various phases and
with different types of the boundaries
as the bulk--edge correspondence 
of the massless Dirac fermion systems. 
\end{abstract}

\maketitle

Massless 
Dirac fermion systems, which are zero gap semiconductors found in 
various situations\cite{Novoselov:2005fk,JPSJ.75.054705,RevModPhys.82.3045,RevModPhys.83.1057,Moore:2009fk}
and characterized by a linear dispersion, are novel semimetallic
materials exhibiting many intriguing phenomena.
A typical
realization of massless Dirac fermion system is celebrated
graphene\cite{Novoselov:2005fk}, which has
been attracted much attention since its
discovery. Not only in conventional solid state materials, but also in
optical lattice systems, the fabrication of massless Dirac fermions
becomes a hot topic very recently\cite{PhysRevLett.103.035301,Tarruell:2012fk}. Among the many unusual properties of the massless 
Dirac fermions, appearance of characteristic edge
states\cite{JPSJ.65.1920,PhysRevLett.89.077002,JPSJ.81.064701} is
important in the view of the 
bulk--edge correspondence, which implies that
 topologically nontrivial bulk states
and appearance of the edge states, i.e., localized modes near the
boundaries, are closely related and reflect each others.
 The concept ``bulk--edge
correspondence'' is established for a topologically nontrivial
gapped state\cite{PhysRevLett.71.3697}. There, a bulk topological number
and number of edge states are connected. Actually, although the massless
Dirac fermion system is gapless at the
Fermi energy and a topological number cannot be well defined, the
bulk--edge correspondence is still at
work\cite{PhysRevLett.89.077002,Hatsugai20091061,PhysRevB.84.195452}.

Instead of the bulk topological number such as the Chern number, the
Berry phase $\theta(k_\parallel)$ plays a central role in the massless
Dirac fermion
systems\cite{PhysRevLett.89.077002,Hatsugai20091061,PhysRevB.84.195452}. 
Here, $\theta(k_\parallel)$ is a bulk quantity parameterized by a
momentum $k_\parallel$, which is parallel to the ``edge''.
Generically the Berry phase $\theta(k_\parallel)$ is gauge dependent
and takes any real number 
in modulo $2\pi$. It is in contrast to the Chern number that is gauge
invariant and 
intrinsically integer\cite{PhysRevLett.49.405}. 
However, with the help of a supplemental symmetry, the Berry phase
is quantized and becomes topological,
that is, adiabatic invariant\cite{PhysRevLett.62.2747,JPSJ.75.123601,Hatsugai10Z2,Hatsugai20091061,JPSJ.82.073708}. 
This is the symmetry protected quantization, which is useful 
in odd dimension. Note that the Chern number and its generalizations
are only well defined in even dimensions. 

The symmetry further plays a crucial role for the topological stability
of the massless Dirac fermions. 
Since the gap closing point has  
co-dimension 3\cite{Berry84,Hatsugai10Z2}, 
the symmetry discussed above is crucial to have a massless Dirac 
fermions in two-dimensions in a generic situation.

As for the 
bulk--edge correspondence of the 
massless Dirac fermions and the stability 
of the doubled  Dirac cones, the chiral symmetry is often
employed\cite{PhysRevLett.89.077002,Hatsugai20091061}. 
 In this paper, with general discrete symmetries,
the idea on the bulk--edge correspondence of 
the massless Dirac fermions and its stability are discussed and 
demonstrated using the fermionic
Shastry-Sutherland (SS) model. This model has not been studied well,
while a {\it spin} model on the SS lattice, which is known as
the orthogonal dimer model, has been extensively
studied following the discovery of the exact ground state wave function\cite{SriramShastry19811069,SHASTRYB,0953-8984-15-9-201,1742-6596-320-1-012019},
and has been realized experimentally\cite{PhysRevLett.82.3168}. 
In the following, we first show that the fermionic
SS model has a rich electronic phase
diagram. Interestingly, the phases with massless Dirac fermions, a quadratic band
crossing point\cite{PhysRevLett.103.046811}, or a pseudospin-1 Weyl
fermion\cite{Dagotto1986383,PhysRevB.85.155451} at the Fermi energy are
accessible by controlling only a few parameters. 
Then the bulk--edge correspondence in 
the fermionic SS model is discussed, focusing on the
phase with massless Dirac fermions. 
Although the fermionic SS model does not respect the chiral symmetry, 
existence of the inversion center or the glide plane play crucial role
in quantization of the Berry phase and the stability of the massless Dirac
fermions. 

\begin{figure}[tb]
 \begin{center}
  \includegraphics[scale=1.0]{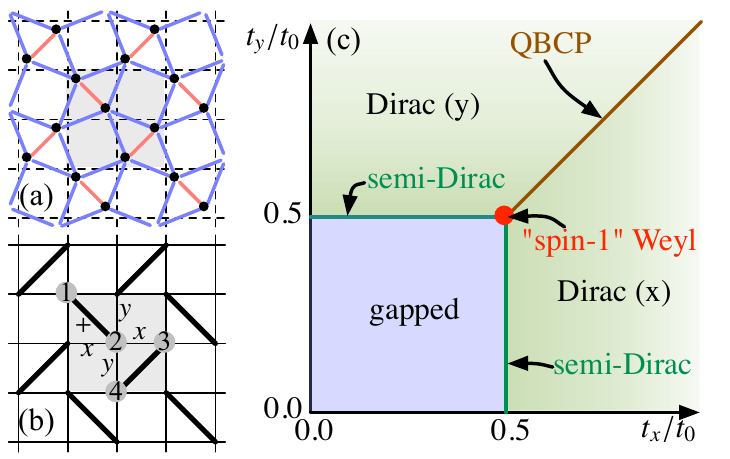}
  \caption{(a) The most ``physical'' Shastry-Sutherland lattice. (b)
  Schematic picture of our model. Bonds named as $+$, $-$, $x$, and
  $y$ are associated with the transfer integrals $t_+$, $t_-$, $t_x$,
  and $t_y$, respectively. Shaded region denotes the unit cell. We give
  numbers one through four to four sites in a unit cell in order to
  distinguish them. (c) Phase diagram for the case of $t_+=t_-=t_0$.} 
  \label{fig1}
 \end{center}
\end{figure}

A possible physical realization of Shastry-Sutherland lattice is
visualized as
Fig.~\ref{fig1}(a). This lattice possesses many symmetries among which
the four-fold rotational symmetry, glide plane symmetry, and inversion
symmetry play particularly important roles in the following arguments. A
simplified picture of the model is shown in Fig.~\ref{fig1}(b). A shaded
region represents a unit cell, which contains four
lattice sites named site 1-4, implying that our model has four bands.
Our Hamiltonian is
\begin{equation}
 H=\sum_{ab}\sum_{\bm{r}\bm{r}'}t_{ab}(\bm{r}-\bm{r}')
  c^\dagger_{\bm{r}a}c_{\bm{r}'b}
  =\sum_{ab\bm{k}}(\hat{H}_{\bm{k}})_{ab}
  c^\dagger_{\bm{k}a}c_{\bm{k}b},
\end{equation}
with
$c_{\bm{k}a}=\frac{1}{\sqrt{N}}\sum_{\bm{r}}\mathrm{e}^{\mathrm{i}\bm{k}\cdot\bm{r}}c_{\bm{r}a}$. 
Here, indices $a$ and $b$ run from 1 thorough 4, representing four
sublattices, while $\bm{r}$ and $\bm{r}'$ represents lattice vectors on
square lattice. 
We employ four parameters $t_+$, $t_-$, $t_x$, and $t_y$ that
correspond to transfer integrals between the sites connected by bonds
indicated as $+$, $-$, $x$, and $y$ in Fig.~\ref{fig1}(b),
respectively. The glide plane symmetry is broken when 
$t_+\neq t_-$, while the four-fold rotational symmetry is broken when 
$t_+\neq t_-$ or $t_x\neq t_y$. In contrast, the inversion symmetry is
always kept with this parameterization. For convenience, we also use
parameters $t_0$,
$t_1$, $\Delta_0$, and $\Delta_1$ defined as $t_\pm=t_0\pm\Delta_0$,
$t_x=t_1+\Delta_1$, and $t_y=t_1-\Delta_1$.
In this study, we neglect spin degrees of freedom, and concentrate on
the half filled case. Namely,
``Fermi energy'' appearing in the following refers to the chemical
potential achieving half filling, and ``gapped state'' means that the
system has a gap between the second and third lowest bands.  

\begin{figure}[tb]
 \begin{center}
  \includegraphics[scale=1.0]{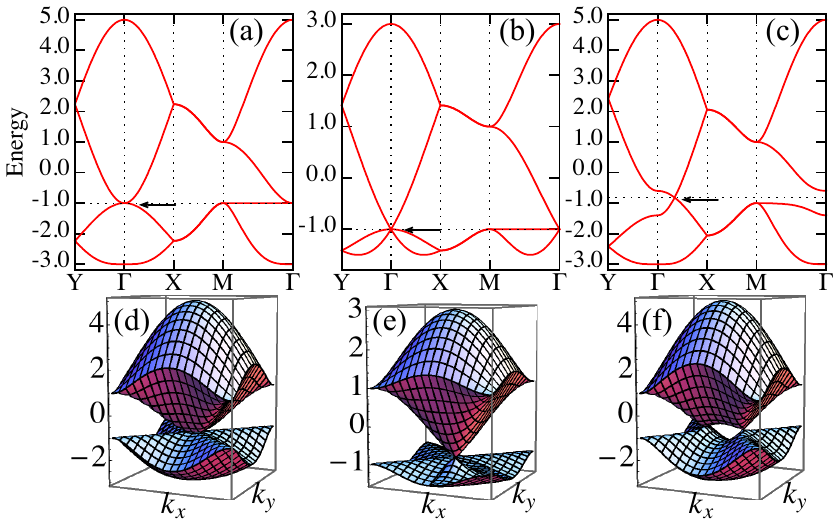}
  \caption{
  Band structures and dispersion relations. $(t_0,\Delta_0)=(1.0,0.0)$
  for the all three cases, while $(t_1,\Delta_1)$ is $(0.0,0.0)$ (a,d),
  $(0.5,0.0)$ (b,e), and $(1.0,0.1)$ (c,f). In (e), a part of the
  dispersion is eliminated so as to make the inside visible.
}
  \label{fig2}
 \end{center}
\end{figure}

The phase diagram obtained for $\Delta_0=0$ ($t_+=t_-$),
which is the case that the two diagonal bonds orthogonal with each other are
equivalent, is shown in Fig.~\ref{fig1}(c). For $t_x=t_y<0.5t_0$, the
system is in a
(trivial) gapped phase. On the other hand, for $t_x=t_y>0.5t_0$, we find
a quadratic band crossing point (QBCP)\cite{PhysRevLett.103.046811}, at
which two parabolic bands,
one is hole-like and the other is electron-like, touch with each other,
at the $\Gamma$-point\cite{SHASTRYB,PhysRevLett.99.227003}. [
Figs.~\ref{fig2}(a) and \ref{fig2}(d).] Note
that the hole-like band is not parabolic in a strict sense in this case,
since it is dispersionless in the $\Gamma$-$M$ direction. QBCP is allowed
to exist if
the system has a four-fold rotational symmetry\cite{Sun:2012fk}, and has
interesting
properties. For instance, the four-fold symmetry can be broken by
electron-electron interaction effects, leading to emergent nematic
phases\cite{PhysRevLett.103.046811}. For $t_x=t_y=0.5t_0$, at which the
transition between the
trivial gapped phase and the phase with QBCP takes place, there exists
``pseudospin-1 Weyl fermion''\cite{PhysRevB.85.155451}, which is
characterized by a linear
dispersion and a three-fold degeneracy\cite{Dagotto1986383}, at the
$\Gamma$-point. [Figs.~\ref{fig2}(b) and \ref{fig2}(e).] 

A finite $\Delta_1$ ($t_x\neq t_y$) imposed in the QBCP phase
immediately leads to a phase with Dirac cones at the Fermi
energy. [Figs.~\ref{fig2}(c) and \ref{fig2}(f).] Namely, two Dirac cones (and
two Dirac points associated with them) are generated as a pair from the
QBCP by a finite $\Delta_1$. The Dirac points are located on
the $k_x$-axis for $\Delta_1>0$, while they are on the $k_y$-axis for
$\Delta_1<0$. Then, if $\Delta_1$ is
continuously modified from positive to negative, the Dirac points first
move on the $k_x$-axis towards the $\Gamma$-point until they merge, and they
next depart from the $\Gamma$-point in the direction of the $k_y$-axis. Note
that the second lowest band is no longer dispersionless on the $\Gamma$-$M$
direction [Fig.~\ref{fig2}(c)], which is important for letting the Dirac
fermions being the only feature appearing at the Fermi energy. If $t_1$
is made smaller and smaller with finite $\Delta_1$, the system
experiences a transition from the phase with Dirac cone to the trivial
gapped phase. The transition between two phases is characterized by an
appearance of a semi-Dirac fermion, whose dispersion is linear in one
direction and parabolic in the other direction. Actually, this type of
disappearance of the Dirac cones is rather general and found in many
other models for Dirac
fermions\cite{springerlink:10.1140/epjb/e2009-00383-0}. 

\begin{figure}[tb]
 \begin{center}
  \includegraphics[scale=1.0]{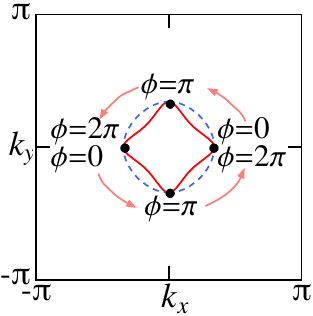}
  \caption{Trajectories of Dirac points for
  $(\Delta_0,\Delta_1)=(\delta_0\sin\phi,\delta_1\cos\phi)$ with
  $t_0=t_1=1.0$. Solid line is for $(\delta_0,\delta_1)=(0.1,0.1)$ while
  dashed line is for $(\delta_0,\delta_1)=(0.2,0.1)$.}
  \label{fig3}
 \end{center}
\end{figure}

When $\Delta_0\neq 0$ simultaneously with 
$\Delta_1\neq 0$, the
Dirac points go into the general points in the Brillouin zone apart
from the high-symmetry lines, i.e., the $k_x$- and $k_y$-axes. In this
case, the symmetry of the system is much lowered, but the inversion (and
time reversal) symmetry is still kept. We will see later that this is
sufficient for stabilizing the massless Dirac fermions by means of the quantized
Berry phase. 
In Fig.~\ref{fig3}, the trajectories of the
Dirac points when $(\Delta_0,\Delta_1)$ is changed according to
$(\Delta_0,\Delta_1)=(\delta_0\sin\phi,\delta_1\cos\phi)$ 
($0\leq\phi\leq 2\pi$) are illustrated for $t_0=t_1=1.0$ and
$(\delta_0,\delta_1)=(0.1,0.1)$ or $(0.2,0.1)$. We find that the
Dirac points wind around the $\Gamma$-point as $\phi$ grows from $0$ to
$2\pi$. Note that the physical state gets back to the original state
after $2\pi$ change in $\phi$, but each Dirac point does not get back to
the original position: two Dirac points interchange their position after
$2\pi$ change in $\phi$. It is also worth noting that once the fermionic
Shastry-Sutherland model is realized in some materials, perturbations
leading to $\Delta_0\neq 0$ and $\Delta_1\neq 0$ can be induced by applying
uniaxial pressure in diagonal or rectangular direction.

\begin{figure}[tb]
 \begin{center}
  \includegraphics[scale=1.0]{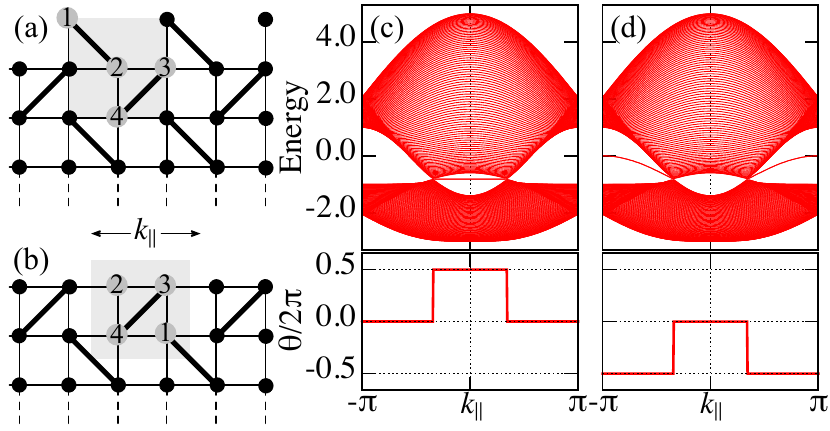}
  \caption{(a,b) Unit cell conventions and edge shapes. (c,d) Edge
  spectra and Berry phase $\theta(k_{\parallel})$
  divided by $2\pi$ as functions of $k_\parallel$ for
  $(t_0,t_1,\Delta_0,\Delta_1)=(1.0,1.0,0.0,0.1)$. (c) is for type 1
  edge (a), while (d) is for type 2 edge (b).}
  \label{fig4}
 \end{center}
\end{figure}

Now, let us discuss the bulk--edge correspondence of the massless Dirac
fermions.
For this purpose, we calculate edge spectra and Berry phase for
fermionic Shastry-Sutherland model. For simplicity, we discuss
the edge parallel to the $x$-axis, but it is possible to extend the
following methods to more general cases\cite{PhysRevB.84.195452}. Edge
spectra are calculated by
making the system with strip (or ribbon) geometries. Here, in order to
make a direct connection to the Berry phase arguments, we set a
rule to make strips for calculation: edges are given by
cutting a periodic system {\it in between the unit cells}. 
With this construction, the edge shapes, or how the system is
terminated at the edge, crucially depend on the convention of the unit
cell since the position of the cut is fixed in between the unit
cells. In this letter, we treat two kinds of unit cell conventions that
lead to two kinds of edge terminations, illustrated as
Fig.~\ref{fig4}(a) and Fig.~\ref{fig4}(b), respectively. Hereafter, we
call the convention in Figs.~\ref{fig4}(a) and \ref{fig4}(b) ``type 1''
and ``type 2'', respectively.
As we limit our attention to the edge parallel to the $x$-axis, Berry
phase\cite{PhysRevLett.62.2747,Hatsugai20091061,PhysRevB.84.195452} is 
defined as
\begin{equation}
 \mathrm{i}\theta(k_{\parallel})=\sum_{n\in\text{filled}}
  \int_{-\pi}^{\pi}\mathrm{d}k_{\perp}
  \langle u_{nk_\parallel k_\perp}|\nabla_{k_\perp}
  |u_{nk_\parallel k_\perp}\rangle,
  \label{Berry_def}
\end{equation}
where $k_\parallel$ and $k_\perp$ are essentially $k_x$ and $k_y$, and
$|u_{nk_\parallel k_\perp}\rangle$ is a Bloch wave function that is a
four-component vector for our four-band tight-binding model. Although we
handle a two-dimensional model here, the extension to $d$ dimensional
cases is straight forward. Namely, we simply regard $k_\parallel$ as a
$d-1$ dimensional vector rather than a number. Actual
evaluation of Eq.~\eqref{Berry_def} is performed using a technique in
Refs.~\onlinecite{PhysRevB.47.1651,PhysRevB.78.054431}.

Calculated edge spectra and $\theta(k_\parallel)/2\pi$ for type 1 and
type 2 conventions with 
$(t_0,t_1,\Delta_0,\Delta_1)=(1.0,1.0,0.0,0.1)$ are plotted as functions
of $k_\parallel$ in Figs.~\ref{fig4}(c) and \ref{fig4}(d). From these
figures, we can extract three important points: (1) appearance of edge
states, (2) quantization of the Berry phase, and (3) an intimate
relation between the edge states and the Berry phase. We explain these
in turn in the following.

Since we chose the parameters so as to have bulk massless Dirac
fermions, bulk
continuum, which is the filled region in
Figs.~\ref{fig4}(c) and \ref{fig4}(d) and contributed from the bulk
states, becomes gapless at the projected Dirac points. 
We find the edge states apart from the bulk continuum
connecting the projected Dirac points for both of the type 1 and 2
cases. The edge states for the type 1 and 2 cases are different due
to the different edge termination. For the type 1 case,
the edge state appears near $k_\parallel$($=k_x$)$=0$, while for the
type 2 case, it appears near $k_\parallel=\pi$. For both
cases, the edge states are dispersive since there is no chiral symmetry
that limits the energy of the edge states, but the edge state shows
almost flat dispersion for the type 1 case.

In general, the quantization of the Berry phase is caused by some
symmetry. In the case of Eq.~\eqref{Berry_def}, it is proven that
the combination of the time reversal and inversion symmetries is
important. These symmetries force $\theta(k_\parallel)$ to obey
\begin{equation}
 \theta(k_\parallel)=-\theta(k_\parallel)+2\pi
  l-2\pi\Delta^I(k_\parallel),
\end{equation}
where
\begin{equation}
  \Delta^{I}(k_\parallel)
  =  \sum_{n\in\text{filled}}
  \frac{1}{2\pi}\int_{-\pi}^{\pi}\mathrm{d}k_\perp
  \mathrm{i}\langle u_{nk_\parallel k_\perp}|\hat{P}_{\bm{k}}^{-1}
  (\partial_{k_\perp}\hat{P}_{\bm{k}})
  |u_{nk_\parallel k_\perp}\rangle
\end{equation}
with $l$ being an integer and $\hat{P}_{\bm{k}}$ being the inversion
symmetry operator satisfying
$\hat{H}_{-\bm{k}}=\hat{P}_{\bm{k}}\hat{H}_{\bm{k}}\hat{P}_{\bm{k}}^{-1}$. Then,
if $\Delta^I(k_\parallel)$ is zero (because $\hat{P}_{\bm{k}}$ has no
$\bm{k}$ dependence, for instance) or some integer (by some symmetrical
reason), $\theta(k_\parallel)$ becomes quantized to $0$ or $\pi$, i.e.,
$Z_2$.
Note that the inversion symmetry alone is sufficient for one-dimensional
models\cite{PhysRevLett.62.2747}, but it must be combined with the time
reversal symmetry for higher dimensional cases. Note also that the 
reflection symmetry whose reflection plane is parallel to the edge alone
can quantize $\theta(k_\parallel)$.  In the case of fermionic SS model,
the glide plane symmetry existing if $t_+=t_-$, plays a role of
the reflection plane symmetry. 

A physical meaning
of the $Z_2$ quantization can be understood from the fact that
$\theta(k_\parallel)$ has close relation to the electronic
polarization\cite{PhysRevB.47.1651}. The inversion or reflection
symmetry gives restrictions for
possible values of the electronic polarization, and these restrictions
appear as the $Z_2$ quantization. However, a special attention is
required in the case that the bulk symmetries are broken after introducing
edges to the system. In our edge construction, edge shapes depend on the
unit cell convention. Then, if we calculate $\theta(k_\parallel)$ using
a unit cell convention that leads to an edge breaking bulk inversion and
glide plane symmetries, $\theta(k_\parallel)$ is not necessarily
quantized even if bulk system without edges has inversion and glide
plane symmetries. This corresponds to the case that
$\Delta^{I}(k_\parallel)$ is noninteger. 

\begin{figure}[btp]
 \begin{center}
  \includegraphics[scale=1.0]{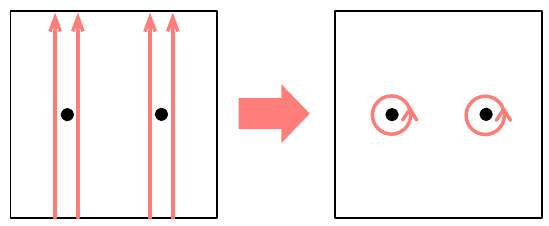}
  \caption{Schematic description of the relation between $\pi$-jumps in
  $\theta(k_\parallel)$ and Dirac points.}\label{fig5}
 \end{center}
\end{figure}

The stability of massless Dirac fermions in two-dimensional systems can be clearly addressed using
the quantized $\theta(k_\parallel)$, which we call $Z_2$ Berry phase. In
order to see
this, we must realize that $\pi$-jump in $\theta(k_\parallel)$ is
directly related to a bulk Dirac fermion. If an 
infinitesimal change in $k_\parallel$, 
$k_\parallel\rightarrow k_\parallel+\delta k$ gives a finite change
between $\theta(k_\parallel)$ and $\theta(k_\parallel+\delta k)$,
the electronic dispersion should have a singularity in the area enclosed
by the two integration paths for $\theta(k_\parallel)$ and
$\theta(k_\parallel+\delta k)$, but, a massless Dirac fermion is nothing more than
a singularity in the electronic dispersion. Furthermore, the value $\pi$
is exactly Berry phase acquired when the integration path encloses a
Dirac point. The idea is described in Fig.~\ref{fig5} as a deformation
of the integration path. Then, as far as the symmetries quantizing
$\theta(k_\parallel)$ are
preserved, massless Dirac fermions are topologically stable, since
$\pi$-jump
cannot be suddenly removed by a small change in parameters when
$\theta(k_\parallel)$ is quantized to 0 or $\pi$: $\pi$-jump
only disappears when two jumps are merged, or parameters themselves are
discontinuously changed. Inversely speaking, if symmetries
preserving $\theta(k_\parallel)$-quantization is broken, massless Dirac fermion
will be no longer stable. In fact, we have checked that when extra terms
breaking the inversion and glide plane symmetries are added to the fermionic
SS model, $\theta(k_\parallel)$ deviates from $0$ or $\pi$, and a gap is
induced at the Dirac point. 

$Z_2$ Berry phase is
also useful in making a criterion for the existence of massless Dirac fermions in a
given model\cite{PhysRev.52.365,PhysRevB.83.245125}. As discussed in
Refs.~\onlinecite{JPSJ.82.034712,JPSJ.82.033703}, there is no need to explore the
entire Brillouin zone to find out Dirac points, thanks to the
$Z_2$ qunatization. Instead, it is
sufficient to check the values of $\theta(k_\parallel)$ at two
$k_\parallel$s, typically at $k_\parallel=0$ and $\pi$. If two
$\theta(k_\parallel)$ take different values, there must be at least one
jump, or equivalently, Dirac point, as far as the quantization is retained. 

The close relation between the appearance of edge states and
$\theta(k_\parallel)$ can be seen in Figs.~\ref{fig4}(c) and \ref{fig4}(d).
Namely, we find edge states for $k_\parallel$ with
$\theta(k_\parallel)=\pi\ \text{mod}\ 2\pi$, while no edge states for
$k_\parallel$ with $\theta(k_\parallel)=0$. Since the $\pi$-jumps are
related to the bulk Dirac points, existence and nonexistence of the edge
states is switched at the Dirac points projected to the edge. Here, we
want to emphasize that, although $\theta(k_\parallel)$ can be calculated
only with bulk information, $\theta(k_\parallel)$ apparently has an
ability to capture the difference in edge terminations, i.e., difference
between type 1 and type 2 edges. This is because
$\theta(k_\parallel)$ does depend on the choice of the basis set since its
definition involves the Bloch wave functions, and different unit cell
conventions are actually connected by a unitary transformation, i.e., a 
transformation of the basis set. In our specific case, $\theta(k_\parallel)$
in type 1 and type 2 conventions are connected as
\begin{equation}
 \theta_{\text{type 2}}(k_\parallel)=\theta_{\text{type 1}}(k_\parallel)-2\pi\rho_1(k_\parallel),\label{theta_trans}
\end{equation}
where $\rho_1(k_\parallel)$ is $k_\parallel$ resolved filling of site 1,
which is explicitly calculated as
\begin{equation}
 \rho_1(k_x)=\sum_{n\in\text{filled}}\frac{1}{2\pi}
  \int_{-\pi}^\pi\mathrm{d}k_y\langle u_{nk_xk_y}|P_1|u_{nk_xk_y}\rangle.
\end{equation}
Here, $P_1$ is a projection operator projecting on the site 1 component. As
far as $t_+=t_-$, $\rho_1(k_\parallel)=0.5$ holds in our model by a
symmetrical reason. Consequently, $\theta_{\text{type 1}}(k_\parallel)$
and $\theta_{\text{ type 2}}(k_\parallel)$ differ by $\pi$. 

Intuitive understanding of this bulk--edge correspondence is possible
with the help of adiabatic continuation when $\theta(k_\parallel)$ is
quantized. We briefly explain
this for the type 2 edge with parameters used in
Fig.~\ref{fig4}(d). Recall that the type 2 edge shows the edge states for
$k_\parallel=\pi$ and no edge
state for $k_\parallel=0$. If $k_\parallel$ is fixed to $\pi$, $\bm{k}$
resolved Hamiltonian $\hat{H}_{\bm{k}}$ can be adiabatically deformed 
{\it without gap closing and keeping $\theta(k_\parallel)$ value} to
the Hamiltonian
corresponding to $t_x=t_y=0$. Then, edge states are readily understood
as dangling states appearing as a result of cutting remained diagonal
bonds for type 2 edge. Importantly, the same adiabatic continuation
cannot be applied to $k_\parallel=0$ case since it leads to the gap
closing, which allows change in quantized $\theta(k_\parallel)$ and
leads to qualitative changes of the system properties. We have to use
different adiabatic continuation, and that continuation should give
Hamiltonian without dangling states for the type 2 edge. 

In summary, we have shown that the fermionic SS model is
a quite important model which hosts many peculiar phases including
the phase with Dirac cones. Since the spin SS model has
been materialized, we believe that it is possible to realize a fermionic
counterpart. Alternatively, the model may be realized in some optical
lattices. Using the SS model, we have also demonstrated roles of the
symmetry for the massless Dirac fermions and the Z$_2$ quantization of
the Berry phase, which also provides the bulk--edge correspondence.

\begin{acknowledgments}
 This work is partly supported by Grants-in-Aid for Scientific Research,
 No.23340112, No.25610101, and No.23540460 from JSPS.
\end{acknowledgments}

\end{document}